\documentclass[conference]{IEEEtran}
\usepackage{graphicx}
\usepackage{subfigure}
\usepackage{amssymb}
\begin{document}



\title{
Effect of interaction strength on the evolution of cooperation
}


\author{\IEEEauthorblockN{Wenfeng Feng}
\IEEEauthorblockA{
School of Computer Science \& Tech-\\nology,Henan Polytechnic University\\
Email: fengwf@hpu.edu.cn}
\and
\IEEEauthorblockN{Yang Li}
\IEEEauthorblockA{
School of Computer Science \& Tech-\\nology,Henan Polytechnic University\\
Email:wuseguang@163.com}
\and
\IEEEauthorblockN{Junhao Yan}
\IEEEauthorblockA{
School of Computer Science \& Tech-\\nology,Henan Polytechnic University}
}
\maketitle
\begin{abstract}
Cooperative behaviors are ubiquitous in nature,which is a puzzle to evolutionary biology,
because the defector always gains more benefit than the cooperator,thus,the cooperator
 should decrease and vanish over time.This typical "prisoners' dilemma" phenomenon  has
 been widely researched in recent years.The interaction strength between cooperators and defectors
 is introduced in this paper(in human society,it can be understood as the tolerance of cooperators).
We find that only when the maximum interaction strength is between two critical values,
the cooperator and defector can coexist,otherwise, 1) if it is greater than the upper value,
the cooperator will vanish, 2) if it is less than the lower value,a bistable state will appear.
\end{abstract}


% \linenumbers

\section{Introduction}

Evolutionary game theory is an efficient way to study evolutionary dynamics,where the fitness of a phenotype is dependent on its frequency relative to other phenotypes in a given population.Different phenotypes compete according to their fitness,the phenotype with greater fitness grow faster and the lesser grow slower.In the end,the competitive phenotype will survive and the weak vanish.
\par
The Prisoner's Dilemma (PD) is a typical model to investigate this problem.In this game,there are two players:a cooperator and a defector.In each round,if the two players are both cooperators,every one will receive R,and if both defectors,they will receive P,a defector can get T by exploiting a cooperator,and the cooperator is left S.In PD,the Condition T\textgreater R\textgreater P\textgreater S should be met.Below is the payoff matrix for this game.
\newline
\begin{table}[h]
\centering
\renewcommand{\arraystretch}{1.2}
\begin{tabular}{p{0.5cm}|p{0.5cm}|p{0.5cm}}

   & C & D \\
   \hline
  C & R & S \\
  \hline
  D & T & P \\
\end{tabular}
\end{table}
\par
Without loss of generality,we assume the size of the population is 1,and the frequency of cooperators is x,defectors y, obviously, x+y=1.The fitness of cooperator and defector is denoted by $f_{c}$ and $f_{d}$ respectively.Thus,in each round:
\par
$$f_{C}=xR+yS\eqno{(1.1a)}$$
$$f_{D}=xT+yP\eqno{(1.1b)}$$
let $\dot{x}=dx/dt,\dot{y}=dy/dt$,so the evolution equations are:
$$\dot{x}=x(f_{C}-\phi)\eqno{(1.2a)}$$
$$\dot{y}=y(f_{D}-\phi)\eqno{(1.2b)}$$
Where $\phi=xf_{C}+yf_{D}$  denotes the mean fitness,above equations are also called the replicator equations\cite{nowak2006evolutionary}.
\par
Because $f_{D}-f_{C}= x(T-R)+y(P-S)>0$,so in the end only the defectors exist in the population,which contradict the real world,since the cooperators always coexist with the defectors in real world.
\par
Cooperation is always vulnerable to exploitation by defectors.Hence,the evolution of cooperation requires specific mechanisms,which allow natural selection to favor cooperation over defection.There had been five mechanisms for the evolution of cooperation\cite{nowak2006five,taylor2007transforming}: direct reciprocity\cite{nowak1992tit,nowak1993strategy}, indirect reciprocity\cite{nowak1998evolution}, kin selection, group selection\cite{traulsen2006evolution}, and network reciprocity \cite{ohtsuki2006simple}(or graph selection).Recently, a mechanism was provided that variable population densities and interaction group sizes can favor the cooperation\cite{wakano2009spatial}.
But,these mechanisms do not take into account that the cooperators could adapt to the change of population frequency and adjust their interaction strength with the defectors. This lead to a natural feedback between population dynamics and interaction strength, and favor the evolution of cooperation.
\begin{figure}[h!]
\centering
\includegraphics[scale=0.6]{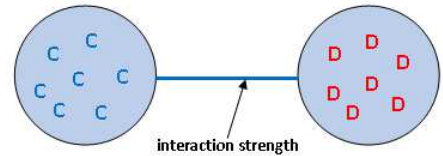}
\caption{Interaction strength.} \label{fig:graph}
\end{figure}
\section{Interaction strength function}
Here,we introduce a function f(x) to represent the variation of interaction strength according to the frequency of cooperators(see Fig.1).It is easy to understand that the cooperators could adjust their interaction strength with the defectors according to the frequency of themselves.Thus,the strength of a cooperator interacts with a defector is f(x),and refuse a defector is 1-f(x).Now,the fitness formula (1.1) are translated into the following formula:
$$f_{C}=xR+yf(x)S\eqno{(2.1a)}$$
$$f_{D}=xf(x)T+yP\eqno{(2.1b)}$$
We study three forms of the interaction strength function f(x):
\par
1) f(x) is a constant.In this situation the cooperator will interact with the defector at a fix strength.Specially,if f(x)=0,cooperators and defectors interact by no means;if f(x)=1,the traditional replicator dynamics is recovered.
\par
2)f(x) is a monotone increasing function about x.In this situation the interaction strength increase along with the frequency of cooperators.Here it can be understood that the more cooperators there are,the more likely they are tolerant,and vice versa(see Fig.2).This conforms to our common sense.
\par
3)f(x) is a monotonic decreasing function about x,this situation is contrary to 2), and it is rarely appear in real world.
\par
\begin{figure}[h!]
\centering
\includegraphics[scale=0.6]{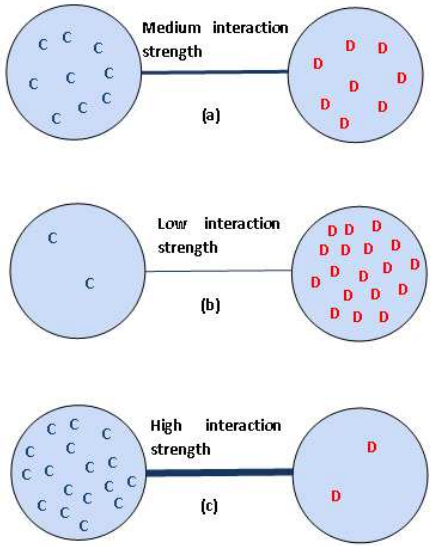}
\caption{Interaction strength when f(x) is a monotone increasing function about x.In this situation the interaction strength increase along with the frequency of cooperators.Here it can be understood that the more cooperators there are,the more likely they are tolerant.} \label{fig:graph}
\end{figure}
In next section we will discuss the situation 1) and 2) respectively,we will show that a fix interaction strength can't give rise to coexist,only when the cooperators adjust their interaction strength in direct proportion to their frequency can lead to coexist.
\section{Population dynamics based on interaction strength}
For simplicity  and without loss of generality,we consider a simpler payoff matrix\cite{wu2010evolution}:
\begin{table}[h!]
\centering
\renewcommand{\arraystretch}{1.2}
\begin{tabular}{p{0.7cm}|p{0.7cm}|p{0.7cm}}

   & C & D \\
   \hline
  C & 1 & 0 \\
  \hline
  D & 1+r & r \\
\end{tabular}
\end{table}
\newline
where r mean how the profitable unilateral defection is,$r\in(0,1)$.
\subsection{f(x) is a constant}
Here,let f(x)=p,p is a constant and $p\in(0,1)$.Now the two players's fitness are as follows:
$$f_{C}=x\eqno{(3.1a)}$$
$$f_{D}=px(1+r)+yr\eqno{(3.1b)}$$
substitute y=1-x into (1.2),we get:
\newline
$$\dot{x}=x(1-x)g(x)\eqno{(3.2a)}$$
where
$$g(x)=(1+r)(1-p)x-r\eqno{(3.2b)}$$
\par
Let $D(x)=\dot{x}$,solve equation D(x)=0,we get three fix points $x^*=1$,0 and $r/[(1+r)(1-p)]$.Below are details about the three situations(see Fig.3):
\par
1)$x^*=0$,here,$D'(x)=-r<0$,so $x^*=0$ is a steady point.\par
2)$x^*=1$,here,$D'(x)=p(1+r)-1$,when $p<1/(1+r)$,it's a steady point, otherwise it's unstable.\par
3)$x^*=r/[(1+r)(1-p)]$,here in order to ensure $x^*\in(0,1)$,the condition must be met: $p<1/(1+r)$.When $p<1/(1+r)$,$D'(x)>0$,so it's a unstable point.
\begin{figure}[h!]
\centering
\includegraphics[scale=0.3]{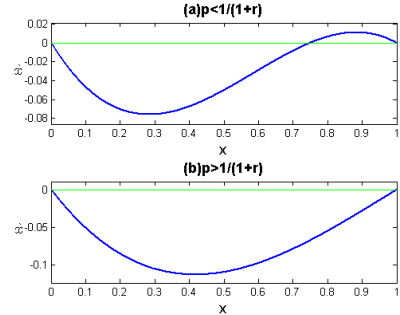}
\caption{The cure of $\dot{x}$ when p take different values.(a)$p<1/(1+r)$,it's  bistable state.(b)$p>1/(1+r)$,it's monostability.We can see no matter what value p takes
,they can't coexist.} \label{fig:graph}
\end{figure}
\par
In conclusion,when f(x) is a constant,cooperators and defectors can not coexist.If $p>1/(1+r)$,the population will be dominated by defectors in the end;if $p<1/(1+r)$,the populations will be homogenous population in the end,as for which dominate the population,it depends on the initial frequency,if $x_{0}< r/[(1+r)(1-p)]$ the defectors win,otherwise the cooperators win(see Fig.4).
\begin{figure}[h!]
\centering
\includegraphics[scale=0.7]{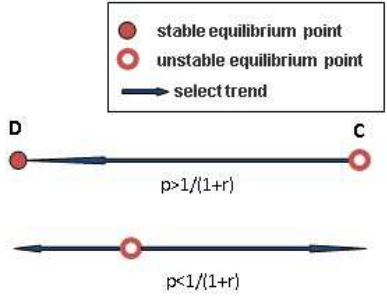}
\caption{Selection Dynamic, corresponding to Fig.3} \label{fig:graph}
\end{figure}
\subsection{f(x) is a monotone increasing function}
Here, f(x) is a monotone increasing function, for simplicity,we let f(x)=kx,where k is a positive constant.Note $f(x)\in[0,k]$,k can be thought  as the max interaction strength or the max tolerance.In this situation,their fitness are as follows:
$$f_{C}=x\eqno{(3.3a)}$$
$$f_{D}=k(1+r)x^{2}+yr\eqno{(3.3b)}$$
substitute y=1-x into (5),we get:
\newline
$$\dot{x}=x(1-x)g(x)\eqno{(3.4a)}$$
where
$$g(x)=-k(1+r)x^{2}+(1+r)x-r\eqno{(3.4b)}$$
Let $D(x)=\dot{x}$,solve equation D(x)=0,we get two boundary  points $x^*=1,0$ and two internal points $x_1,x_2$,they are two roots of equation g(x)=0.
Below are details about these points.
\subsubsection{dynamics on boundary  points}
Below is the dynamics on points $x^*=0$ and $x^*=1$.\par
	$x^*=0$, In this situation, $D'(0)=g(0)=-r<0$,so $x^*=0$ is a unstable point.\par
	$x^*=1$, In this situation, $D'(1)=-g(1)=k+rk-1$,when $k<1/(1+r)$,it's a steady point,otherwise,it's a unstable point.
\subsubsection{dynamics on internal  points}
In this situation,there are two fix points denoted by $x_1,x_2$,they are two roots of the parabolic equation g(x)=0,suppose $x_1 \leqslant x_2$.
At first, To ensure the existence of the solution, $\bigtriangleup=(1+r)(1+r-4rk)\geqslant0$ must be satisfyed,that is,
$$k\leqslant0.25(1+1/r)\eqno{(3.5)}$$
\par
Note $x_1x_2= k(1+r)/r>0$ and $x_1+x_2=1/k>0$,we know $x_1>0,x_2>0$.Note $D'(x^*)=(1-x^*)x^*g'(x^*)$,so to determine the stability of the two points,
it's only necessary to consider the situation of $g'(x^*)$.Below is the discussion of the stability of the two point(see Fig.5).
\par
1) $x_1<1,x_2<1$,to ensure this,$g(1)<0$ and $(x_1+x_2)/2<1$ must be satisfied,that is,$k>1/(1+r)$must be satisfied.In this situation,
we get $g'(x_1)>0$ and $g'(x_2)<0$,so $x_1$ is unstable and $x_2$ is stable.
\par
2)$x_1<1,x_2>1$,to ensure this,$g(1)>0$ and $(x_1+x_2)/2<1$ must be satisfied,that is,$k<1/(1+r)$ must be satisfied.In this situation,
there is only one internal point $x_1$,and it is unstable.
\par
3)$x_1>1,x_2>1$,to ensure this,$g(1)<0$ and $(x_1+x_2)/2>1$ must be satisfied,that is,$k>1/(1+r)$ and $k<0.5$ must be satisfied,because $1/(1+r)<0.5$,
so this condition can never be met, so there is at least one internal point.
\subsubsection{Summary}
By $\S3.1$ and $\S3.2$,let $k_1=1/(1+r),k_2=0.25(1+1/r)$, $x_1$ and $x_2$ are the smaller root and larger root of equation g(x)=0 respectively.The conclusions are drawn as follows:
\par
1)	If $k<k_1$,x=1 and x=0 are bistable state,the demarcation point is $x_1$.\par
2)	If $k_1<k<k_2$,x=0 and $x=x_2$ are stable points,$x=x_1$ is unstable point.The cooperators and defectors can
coexist on the point $x=x_2$,but when $x<x_1$,the cooperators will still become extinct.\par
3)	If $k>k_2$,there is no internal point,x=0 is the global stable point,and x=1 is the global unstable point,in the end,only the defectors exist in the population.
\begin{figure}[h!]
\includegraphics[scale=0.4]{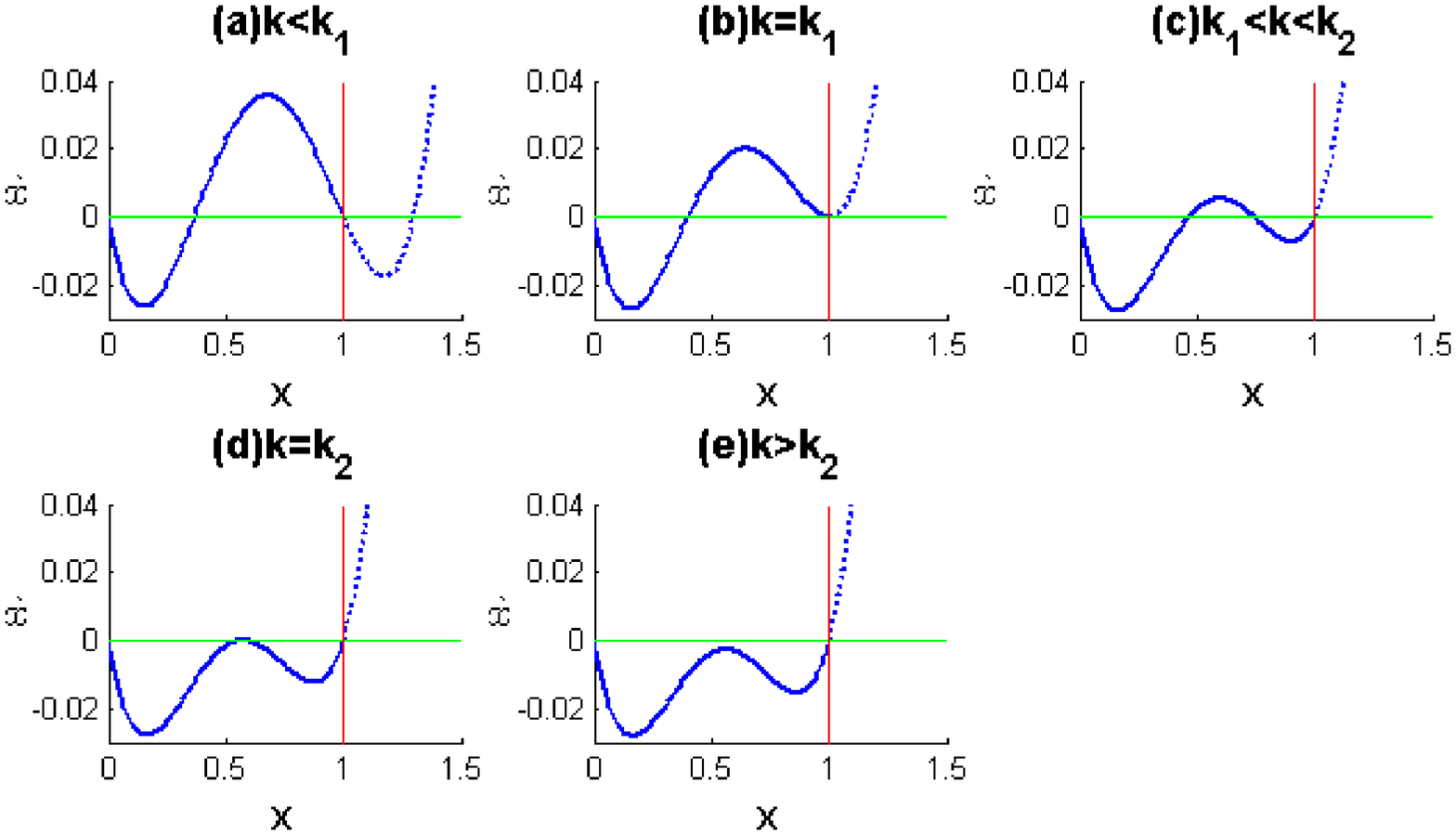}
\caption{The cure of $\dot{x}$ when k takes different values.} \label{fig:graph}
\end{figure}
\begin{figure}[h!]
\centering
\includegraphics[scale=0.75]{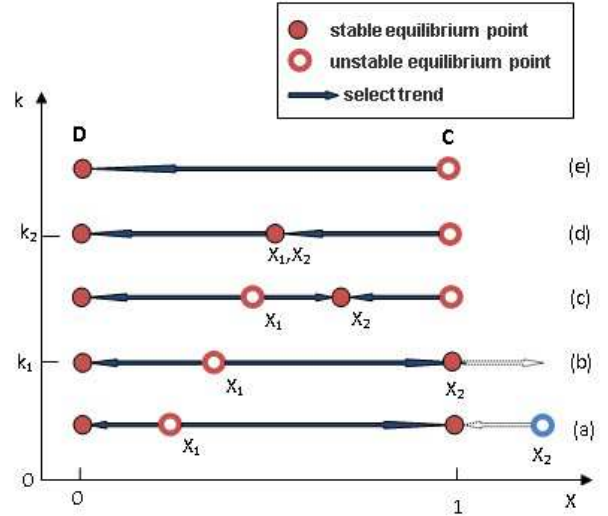}
\caption{Selection Dynamic, corresponding to Fig.5} \label{fig:graph}
\end{figure}
\subsubsection{Numerical simulation}
Here,we simulate this model with computer,where f(x)=kx,r=0.2.At t=0,there are 50 original  states say $\bold{x}=\{x_1,x_2\cdots x_{50}\}$,each is a random float in (0,1).In each time step,
$$x(t)=x(t-1)+ \dot{x}\times step$$
Where step mean step size,here step=1.The result is shown in Fig.7.
\begin{figure}[h!]
\centering
\includegraphics[scale=0.45]{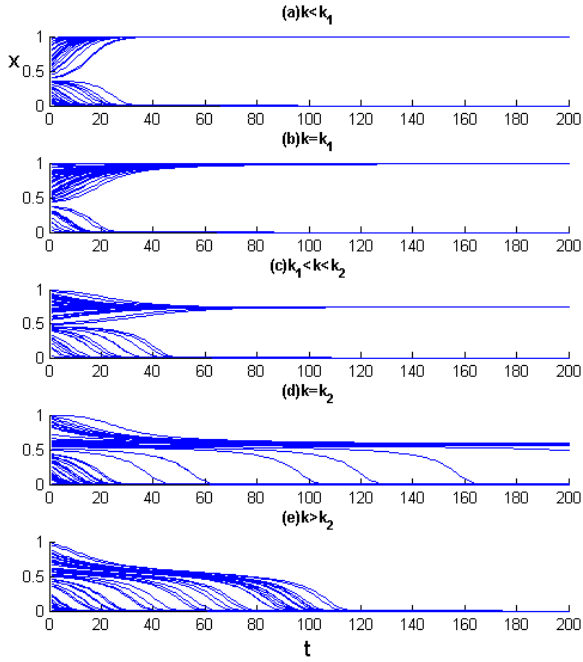}
\caption{Numerical simulation on f(x)=kx,r=0.2. corresponding to Fig.5.Note (d) is a semi-stable point,that means on this point,the states will decay to the point x=0
in a "Half-Life" form.} \label{fig:graph}
\end{figure}
\section{Conclusion}
The Prisoner's Dilemma is a general metaphor for the problem of cooperation\cite{taylor2007evolution,stevens2004nice,turner1999prisoner},in this paper,we introduce the interaction strength,which leads to coexist of cooperators and defectors.
\par
It's shown that when interaction strength is constant,it can not leads to coexist.This suggests that it's unwise for the cooperators to be fixed and unchangeable to the defectors.
\par
When the interaction strength is a monotone increasing function about the frequency of cooperators,there is a max interaction strength.Only when the max interaction strength is between two critical value,the cooperator and defector can coexist.The two critical points is 1/(1+r) and (1+r)/(4r) denoted by $k_1$ and $k_2$  respectively.Note that 1+r is the payoff of a defector to plunder a cooperator,r is the payoff of a defector meet a defector,1 is the payoff of a cooperator meeting a cooperator.So $k_1$ can be understood as the ratio of the payoff of a cooperator to the payoff of a defector when they meet the same cooperator,the same to $k_2$.
\par
In general PD payoff matrix with b and c as follows:
\begin{table}[h!]
\centering
\renewcommand{\arraystretch}{1.2}
\begin{tabular}{p{0.7cm}|p{0.7cm}|p{0.7cm}}

   & C & D \\
   \hline
  C & b-c & -c \\
  \hline
  D & b & 0 \\
\end{tabular}
\end{table}
\newline
where a cooperator cost c to give his partner b.Note,r correspond to c/b,1 corresponds to b,we can get the following formula:
$$\frac{b}{b+c}<k<\frac{b+c}{4c}$$
where k means the max interaction strength or the max tolerability.In social life,the cooperator can be seen as producer,and the defector predator.From the above formula we can see
if the max tolerability is in between b/(b+c) and (b+c)/(4c) the both can coexist,which is a general phenomenon in most society,if it's greater than (b+c)/(4c),it will leads to social instability,
if it's smaller than b/(b+c),it will towards a society in which almost everyone is a cooperator.
%% The Appendices part is started with the command \appendix;\


%% References

\section{Acknowledgments}
The authors gratefully acknowledge the editors and the anonymous referees' delightful suggestions. This work was supported by National Nature Science Foundation of China No. 60703053.
\bibliographystyle{plain}
\bibliography{mywenxian}






% \end{thebibliography}

\end{document}